\documentclass[12pt,epsfig]{article}
\usepackage{graphicx}
\usepackage{array}
\usepackage{comment}
\usepackage{mathrsfs}
\usepackage{amsfonts}
\usepackage{mathtools}
\usepackage{bbold}
\usepackage{bbm}
\usepackage{amsmath,amssymb}
\usepackage[latin1]{inputenc}

\begin{document}

\title{{\bf Non-linear Electrodynamics \\ derived from the Kaluza-Klein Theory}}

\maketitle

\author{R. Kerner, LPTMC, Tour 23-13, 5-e \'etage, Bo\^{i}te 121, Sorbonne-Universite, 4, Place Jussieu, 75252 Paris Cedex 05, France,~~E-mail: richard.kerner@sorbonne-universite.fr} 

\vskip 0.5cm
\indent
\hskip 0.7cm
{\bf Abstract}
\vskip 0.2cm
\indent
The lagrangian of the Kaluza-Klein theory, in its simplest five-dimensional version, should include not only the scalar curvature $R$,
but also the quadratic Gauss-Bonnet invariant. The general lagrangian is computed and the resulting non-linear equations which generalize 
Maxwell's system in a quite unique way are investigated. The possibility of the existence of static solutions is presented, and the qualitative 
behaviour of such solutions is discussed.

\newpage

\section{Introduction}

Although we know that the five-dimensional Kaluza-Klein theory  [1], [2] is not viable, one of the reasons being the
absence of chiral spinors in five space-time dimensions, it is present in some way or another in a more realistic theory as a restriction
to the $U(1)$-subgroup of the full structural group. As such, it should describe fairly well the equations governing the electromagnetic sector
of any unified theory when other interactions can be neglected.

The classical Maxwell equations are known to describe very well the behaviour of the electromagnetic field in the situations when 
the quantum effects can be put aside, i.e. when the fields are not strong enough. The quantum effects, on their side, perturb the purely linear 
behaviour of the Maxwellian field, giving rise to the non-linear corrections in form of series in powers of $\alpha = 1/137$. It is not absurd 
to think that the perturbations would get together to yield some simple non-linear equations for the mean values of the fields, i.e. that the
quantum electrodynamics might be replaced, at least in some situations, e.g. static configurations, by some version of non-linear electrodynamics.

In $1912$ Mie [3], and later, in $1924$, Born and Infeld [4] proposed two versions of the non-linear electrodynamics, hoping to explain the existence 
of apparently non-singular static configurations such as electrons. Their results were not convincing, mainly because the symmetry between the electric 
and magnetic fields was broken from the beginning. Roughly speaking, it seems improbable to obtain a configuration with only electric field present,
putting ${\bf B} = 0$ everywhere. The lagrangians they have chosen depended quite arbitrarily on two invariants of the electromagnetic tensor, 
$F_{\mu \nu} F^{\mu \nu} = - 2 ( {\bf E}^2 - {\bf B}^2)$ and $F_{\mu \nu}{^{*}}F^{\mu \nu} = {\bf E} \cdot {\bf B}$.
The second invariant not appearing in the linear theory has been also discarded in the non-linear version of Mie.

An infinite number of theories based on the lagrangians depending on $F_{\mu \nu} F^{\mu \nu}$ and  $(F_{\mu \nu}{^{*}}F^{\mu \nu})^2$ (the square is needed
in order to keep the invariance of the theory under space reflections) can be produced if we lack any guiding principle that would fix the form
of the lagrangian. Nowadays such principle can be found from the geometrical invariants in many dimensions whose form is suggested by the string
theories [5], [6]. Let us remind that the Kaluza-Klein theory, as well as its improvements by Jordan [7] and Thiry [8] has been based on the 
Einstein-Hilbert variational principle in five-dimensional space, with the total lagrangian equal to the scalar curvature of the metric. This
lagrangian is unique in four dimensions, because already the second invariant of the Riemann tensor,
\begin{equation}
I_2 = R_{ABCD}R^{ABCD} - 4 \; R_{AB} R^{AB} + R^2
\label{GaussBonnet}
\end{equation}
turns out to be a pure divergence and does not modify the equations of motion. The invariant (\ref{GaussBonnet}) is the only quadratic combination
of the Riemann tensor leading under variation to the second-order equations (see Madore [9], Lanczos [10], M\"uller-Hoissen [11]). In five dimensions,
however, this invariant is no more a divergence, therefore there is no reason to not include it in the full theory. As we shall see, this fixes
the lagrangian in five dimensions, leaving the place for the arbitrariness only in the choice of one dimensional parameter. 
 This will be the starting point for the non-linear modification of the electrodynamics. In our calculus we shall deliberately discard the
gravitational and scalar fields, in principle also present, but supposedly too weak to influence the behaviour of the electromagnetic field 
at short distances.

\section{The derivation of the modified Maxwell's system }

The well-known ansatz for the metric tensor in five-dimensional Kaluza-Klein theory reads as follows:

\begin{equation}
g_{AB} = \begin{pmatrix} g_{\mu \nu} +  A_{\mu} A_{\nu} & A_{\mu} \cr
 A_{\nu} & 1 \end{pmatrix}, 
\label{KKmetricgen1}
\end{equation}
where $A,B=0,1,2,3,5$ and $\mu,\nu=0,1,2,3$ 

We shall deliberately discard the scalar field $g_{55} (x^{mu})$, putting $g_{55} = 1$, although the full system will be overconstrained. 
We shall also put $g_{mu \nu} = \eta_{\mu \nu} =  diag (+,-,-,-)$, saving only the electromagnetic part of the lagrangian.

It is quite easy to compute all the non-vanishing components of the Riemann tensor $R_{AB}$ of the metric (\ref{KKmetricgen1}),
especially in the non-holonomous frame

\begin{equation}
\xi_{\mu} = \partial_{\mu} - A_{\mu} \partial_5, \; \; \; \xi_5 = \partial_5.
\label{nonhframe}
\end{equation}
They are as follows (cf. [12], \; [13]) :
\begin{equation}
R_{\mu \nu \lambda \rho} = \frac{1}{4} \; F_{\mu \lambda} F_{\rho \nu} - F_{\nu \lambda} F_{\rho \mu} + 2 \; F_{\mu \nu} F_{\rho \lambda},
\label{Riemann44}
\end{equation}
\begin{equation}
R_{\mu 5 \lambda \rho} = \frac{1}{2} \; \partial_{\mu} F_{\rho \lambda}
\label{Riemann45}
\end{equation}
\begin{equation}
R_{\mu 5 5 \lambda} = \frac{1}{4} \; \eta^{\nu \rho} \; F_{\mu \lambda} F_{\rho \nu}.
\label{Riemann55}
\end{equation} 
The Ricci tensor is then
\begin{equation}
R_{\mu \nu} = \frac{1}{2} \; \eta^{\lambda \rho} F_{\mu \lambda} F_{\nu \rho}
\label{Ricci44}
\end{equation}
\begin{equation}
R_{\mu 5} = \frac{1}{2} \; \eta^{\nu \rho} \partial_{\nu} F_{\rho \mu} 
\label{Ricci45}
\end{equation}
\begin{equation}
R_{55} = \frac{1}{4} \; \eta^{\mu \nu} \eta^{\lambda \rho} \; F_{\mu \lambda} F_{\rho \nu}
\label{Ricci55}
\end{equation}
and the scalar curvature
\begin{equation}
R = - \frac{1}{4} \eta^{\mu \lambda} \eta^{\nu \rho} \;  F_{\mu \nu} F_{\lambda \rho}
\label{Rscalar}
\end{equation}
with
\begin{equation}
F_{\mu \nu} = \partial_{\mu} A_{\nu} - \partial_{\nu} A_{\mu}.
\label{Fdef}
\end{equation}
The last expression (\ref{Rscalar}) is the lagrangian of Maxwell's theory. Let us introduce a more general variational principle, 
based on the lagrangian density
\begin{equation}
{\cal{L}} = (e^2 R + I_2) \; \sqrt{\mid g \mid}
\label{R+I2}
\end{equation}
with the dimensional parameter $e^2$ to be determined later. $e^2$ has the dimension of ${\rm{cm}}^{-2}$ in order to make the whole
expression homogeneous, as dim [R] = cm$^{-2}$ and dim $I_2 = $ cm$^{-4}$.

The invariant $I_2$ for the metric (\ref{KKmetricgen1}) is easily calculated and is found to be (discarding the pure divergence term
equal to $\partial_{\mu} (F_{\rho \lambda} \partial^{\mu} F^{\rho \lambda}) - 2 \partial^{\nu} (F_{\rho \lambda} \partial_{\nu} F^{\mu \lambda})$:
\begin{equation}
I_2 = \frac{3}{16} \; \left[(F_{\mu \nu} F^{\mu \nu})^2 - 2 F_{\mu \lambda} F_{\nu \rho} F^{\mu \nu} F^{\lambda \rho} \right].
\label{InvI2}
\end{equation} 
For fixed Minkowskian metric $\\eta_{\mu \nu}$ we can put $\sqrt{\mid g \mid} = 1$ and write the full lagragian as
\begin{equation}
{\cal{L}} = -\frac{1}{4} \; F_{\mu \nu}F^{\mu \nu} + 
\frac{3 \varepsilon}{16 e^2} \left[ F_{\mu \nu} F^{\mu \nu})^2 - 2 F_{\mu \lambda} F_{\nu \rho} F^{\mu \nu} F^{\lambda \rho} \right],
\label{Inv1Inv2}
\end{equation}
with $\varepsilon$ a numerical parameter to be determined.

The equations of motion in vacuo are then
\begin{equation}
\partial_{\lambda} \left[ F^{\lambda \rho} - \frac{3 \varepsilon}{16 e^2} (F_{\mu \nu} F^{\mu \nu}) F^{\lambda \rho}
+ \frac{3 \varepsilon}{e^2} F_{\mu \nu} F^{\lambda \mu} F^{\rho \nu} \right].
\label{Eqsmo1}
\end{equation} 
The identities
\begin{equation}
\partial_{\mu}F_{\lambda \rho} + \partial_{\lambda}F_{\rho \mu} + \partial_{\rho} F_{\mu \lambda} =0
\label{BianchiF}
\end{equation}
hold by definition (\ref{Fdef}), too.

Both the lagrangian and the equations become much more transparent when expressed by means of the fields {\bf E} and {\bf B},
{\bf D} and {\bf H}:
\begin{equation}
{\cal{L}} = \frac{1}{2} ({\bf E}^2 - {\bf B}^2) + \frac{3 \varepsilon}{2 e^2} ({\bf E} \cdot {\bf B})^2.
\label{LwithEB}
\end{equation}
As we see, the new term contains only the square of the second invariant of the electromagnetic field.
he full set of modified Maxwell's equations is:
$$ {\rm div}\;  {\bf B} = 0, \; \; \; \; \; \; {\bf rot}\; {\bf E} = - \frac{\partial {\bf B}}{\partial t}, $$
\begin{equation}
{\rm div} \; {\bf D} = - \frac{3 \varepsilon}{e^2} \; {\bf B} \cdot {\bf grad} ({\bf E} \cdot {\bf B}), 
\; \; \; {\bf rot} \; {\bf H} = \frac{\partial {\bf D}}{\partial t} + \frac{3 \varepsilon}{e^2} 
\left[  {\bf H} \frac{\partial ({\bf E} \cdot {\bf B})}{\partial t} - {\bf E} \times {\bf grad} ({\bf E} \cdot {\bf B}) \right].
\label{modMaxwell1}
\end{equation}
In what follows we shall use the units in which $c =1$, and in which we can put in the vacuum ${\bf E} = {\bf D}$ and ${\bf H} = {\bf B}$.
Therefore the equations in vacuum will be
$$ {\rm div}\;  {\bf B} = 0, \; \; \; \; \; \; {\bf rot}\; {\bf E} = - \frac{\partial {\bf B}}{\partial t}, $$
\begin{equation}
{\rm div} \; {\bf E} = - \frac{3 \varepsilon}{e^2} \; {\bf B} \cdot {\bf grad} ({\bf E} \cdot {\bf B}), 
\; \; \; {\bf rot} \; {\bf B} = \frac{\partial {\bf E}}{\partial t} + \frac{3 \varepsilon}{e^2} 
\left[  {\bf B} \frac{\partial ({\bf E} \cdot {\bf B})}{\partial t} - {\bf E} \times {\bf grad} ({\bf E} \cdot {\bf B}) \right].
\label{modMaxwell2}
\end{equation}
When $\varepsilon$ is put equal to zero, the equations recover their usual Maxwellian form. Two other possibilities, up to a scale
that can be incorporated in $e^2$, are $\varepsilon = +1$ or $-1$.

\section{Discussion of general properties of equations}

The non-homogeneous couple of equations,
\begin{equation}
{\rm div} \; {\bf E} = - \frac{3 \varepsilon}{e^2} \; {\bf B} \cdot {\bf grad} ({\bf E} \cdot {\bf B})
\label{Maxmod1}
\end{equation}
and
\begin{equation}
{\bf rot} \; {\bf B} = \frac{\partial {\bf E}}{\partial t} + \frac{3 \varepsilon}{e^2} 
\left[  {\bf B} \frac{\partial ({\bf E} \cdot {\bf B})}{\partial t} - {\bf E} \times {\bf grad} ({\bf E} \cdot {\bf B}) \right]
\label{Maxmod2}
\end{equation}
can be implemented by adding the charge density $\rho$ to the right-hand side of (\ref{Maxmod1}) and the current density ${\bf j}$ 
to the right-hand side of  (\ref{Maxmod2}). However, even in the absence of these ``external sources'', the right-hand sides of the
eqs. (\ref{Maxmod1}) and [\ref{Maxmod2}) behave like conserved induced charge and current densities; their conservation is totally
independentof eventual other non-induced similar objects. As a matter of fact, let us compare:
$$\frac{\partial}{\partial t} ( {\rm div} \; {\bf E}) = - \frac{3 \varepsilon}{e^2}  \frac{\partial {\bf B}}{\partial t} \cdot  
{\bf grad} ({\bf E} \cdot {\bf B}) - \frac{3 \varepsilon}{e^2} {\bf B} \cdot {\bf grad} \frac{\partial ({\bf E} \cdot {\bf B})}{\partial t} = $$
\begin{equation}
= - \frac{3 \varepsilon}{e^2} ( {\bf rot} {\bf E}) \cdot {\bf grad} ({\bf E} \cdot {\bf B}) -\frac{3 \varepsilon}{e^2} {\bf B} \cdot {\bf grad} 
\frac{\partial ({\bf E} \cdot {\bf B})}{\partial t}
\label{ddivEdt}
\end{equation}
and 
$$ {\rm div} \; \frac{\partial {\bf E}}{\partial t} = {\rm div} ({\bf rot} {\bf B})
- \frac{3 \varepsilon}{e^2} {\rm div} \left( {\bf B} \; \frac{\partial ({\bf E} \cdot {\bf B})}{\partial t} \right) 
- \frac{3 \varepsilon}{e^2} {\rm div} ( {\bf E} \times {\bf grad} ({\bf E} \cdot {\bf B})) = $$
\begin{equation}
= \frac{3 \varepsilon}{e^2} ( {\rm div} {\bf B})  \frac{\partial ({\bf E} \cdot {\bf B})}{\partial t}
-  \frac{3 \varepsilon}{e^2} {\bf B} \cdot {\bf grad} \frac{\partial ({\bf E} \cdot {\bf B})}{\partial t}
+ \frac{3 \varepsilon}{e^2} ({\bf rot} {\bf E}) \cdot {\bf grad} ({\bf E} \cdot {\bf B}).
\label{divdEdt}
\end{equation}
because
$${\rm div} {\bf B} = 0, \; \; \; {\bf rot}({\bf grad} f ) = 0, \; \; \; \; 
{\rm div} ( {\bf a} \times {\bf b} ) = {\bf b} \cdot( {\bf rot} {\bf a}) - {\bf a} \cdot( {\bf rot} {\bf b}), $$ 
therefore
\begin{equation}
\frac{\partial }{\partial t} \left[ - \frac{3 \varepsilon}{e^2}  {\bf B} \cdot {\bf grad} ( {\bf E} \cdot {\bf B}) \right]
+ {\rm div}  \left[ \frac{3 \varepsilon}{e^2} {\bf B} \frac{\partial ({\bf E} \cdot {\bf B})}{\partial t} 
- {\bf E } \times {\bf grad} ({\bf E} \cdot {\bf B}) \right] = 0.
\label{conserv1}
\end{equation}
We shall denote the induced charge density by ${\rho}_{ind}$:
\begin{equation}
{\rho}_{ind} =  - \frac{3 \varepsilon}{e^2}  {\bf B} \cdot {\bf grad} ( {\bf E} \cdot {\bf B}),
\label{rhoind}
\end{equation}
and the induced current density by ${\bf j}_{ind}$:
\begin{equation}
{\bf j}_{ind} = \frac{3 \varepsilon}{e^2} {\bf B} \frac{\partial ({\bf E} \cdot {\bf B})}{\partial t} 
- {\bf E } \times {\bf grad} ({\bf E} \cdot {\bf B}) 
\label{jind}
\end{equation}
with 
\begin{equation}
\frac{\partial {\rho}_{ind}}{\partial t} + {\rm div} ({\bf j}_{ind}) = 0.
\label{conserv2}
\end{equation}
The theory will not need any non-induced charges if we can prove the existence of charged stable static solutions,
localized in space (solitons).

If we form the sum
\begin{equation}
{\bf B} \cdot \frac{\partial {\bf B}}{\partial t}  + {\bf E} \cdot \frac{\partial {\bf E}}{\partial t}  
\label{sumBEdt}
\end{equation}
we shall easily find another conservation law:
\begin{equation}
\frac{\partial }{\partial t} \left[ \frac{1}{2} ( {\bf E}^2 + {\bf B}^2 ) + \frac{3 \varepsilon}{e^2} ({\bf E} \cdot {\bf B})^2 \right]
= {\rm div} \; ({\bf E} \times {\bf B})
\label{Econserve}
\end{equation}
Thus the Poynting vector in this theory is the same as in the linear electrodynamics, whereas the energy density contains a new term,
as compared with the classical theory:
\begin{equation}
{\cal{H}} = \frac{1}{2} ( {\bf E}^2 + {\bf B}^2 ) + \frac{3 \varepsilon}{e^2} ({\bf E} \cdot {\bf B})^2
\label{Energy}
\end{equation}
Note that the parameter $\varepsilon$ has to be positive, in order to ensure the positivity of the energy. From now on we shall
set $\varepsilon = 1$, leaving only the coupling constant $e^2$ to be determined.
\vskip 0.2cm
The hamiltonian  ${\cal{H}}$ can be also found from the lagrangian ${\cal{L}}$ directly. The only time derivatives present in ${\cal{L}}$
are $\partial_0 A_k, \; k=1,2,3$ which enter through the combination $E_k = \partial_0 A_k - \partial_k A_0$. Therefore $E_k$, the components
of the electric field, can be chosen as generalized velocities, so that
\begin{equation}
{\cal{H}} = {\bf E} \cdot \frac{\partial {\cal{L}}}{\partial {\bf E}} - {\cal{L}}.
\label{Hcanonic} 
\end{equation}
which yields the same result.
\vskip 0.2cm
The canonical quantization will be much more complicated than in usual theory. Although the matrix
\begin{equation}
\frac{\partial^2 {\cal{L}}}{\partial E_k \partial E_l} = \delta^{kl} + \frac{3}{e^2} B^k B^l
\label{hessian}
\end{equation} 
is obviously positive definite, and therefore always invertible, its inverse, which is necessary to express $E_k$ in terms of the canonical momenta 
$\pi^k = \partial {\cal{L}}/\partial E_k$, is an infinite series of even powers of $B^k$. This poses a serious problem, in additional to the
usual constraint problem $\partial_0 A_0 = 0$. 

Whenever the fields ${\bf E}$ and ${\bf B}$ are orthogonal to each other, our system in vacuum (\ref{Maxmod1}, \ref{Maxmod2}) 
coincides with Maxwell's equations. Such is the case of the electromagetic waves, which are also solutions to the equations (\ref{Maxmod1}, \ref{Maxmod2}).
Moreover, these solutions are stable with respect to perturbations. As a matter of fact, any deviation from the usual solution in which
${\bf E}$ is everywhere orthogonal to ${\bf B}$, leads automatically to the rise in the energy ${\cal{H}}$, which ensures the stability.

Let us also note that the energy-momentum tensor could be obtained directly as
\begin{equation}
T_{\mu \nu} = \frac{\partial ( \sqrt{\mid g \mid} {\cal{L}}}{\partial g^{\mu \nu}}
\label{enmomtensor}
\end{equation}
yielding the same expressions for the Hamiltonian $T_{00}$ and the Poynting vector $P_k = T_{0k}$.

\section{Search for static non-singular solutions}

Let us rewrite the equations (\ref{Maxmod1} and ({\ref{Maxmod2}) in the stationary case, when all the time derivatives vanish:
$${\rm div} \; {\bf B} = 0, \; \; \; \; \; {\bf rot} \; {\bf E} = 0,$$
\begin{equation}
{\rm div} \; {\bf E} = - \frac{3}{e^2} {\bf B} \cdot {\bf grad} ({\bf E} \cdot {\bf B}), \; \; \; \; 
{\bf rot} {\bf B} = - \frac{3}{e^2} {\bf E} \times {\bf grad} ({\bf E} \cdot {\bf B} ).
\label{Maxstat}
\end{equation}
It would be very interesting to obtain a static and non-singular solution of this system, having a finite energy and behaving like a soliton.

This is excluded in the linear case, therefore, if such solution exists, both fields ${\bf E}$ and ${\bf B}$ must be different from
zero and non-orthogonal at least in some finite domain of space. We should also impose the rapid enough vanishing of both fields at infinity.
\vskip 0.2cm
The spherical symmetry for ${\bf B}$ leads immediately to the singularity at the origin; if the condition ${\rm div} {\bf B}$ is 
to be maintained everywhere, then the lines of force of the field ${\bf B}$ have to be closed. The lines of the local current ${\bf rot} {\bf B}$
must be closed, too; this suggests the axial symmetry in which the current would have only the azimuthal component, and the field ${\bf B}$ would be 
everywhere perpendicular to the azimuthal unit vector ${\bf e}_{\varphi}$ (in cylindrical coordinates $(\rho = \sqrt{x^2 + y^2)}, \; z, \; \varphi) $), 
i.e. ${\bf B}$ having its components along ${\bf e}_{z}$ and ${\bf e}_{\rho}$ only. Also the field ${\bf E}$ should have only the $z$ and $\rho$
components; then the Poynting vector ${\bf P} = {\bf E} \times {\bf B}$ will have the azimuthal component only.

Although we shall see that such a configuration can not be obtained without singularity, it has some remarkable symmetry properties
which we will discuss here.
\vskip 0.2cm
The trilinear combinations on the right-hand sides of equations (\ref{Maxstat}) produce the induced charge and current densities. 
The current having only azimuthal component everywhere will produce magnetic field which at great distances is similar to that of 
a circular distribution of currents, and will be that of a magnetic dipole. At the same time, one can expect a non-vanishing charge
concentration falling off quite rapidly with distance from the origin, at large distances ${\bf E}$ should be then similar to the electric 
field of a concentrated charge.
\vskip 0.2cm
All these conditions put together lead to the following symmetry properties of the components ${\bf E}$ and ${\bf B}$:
\begin{equation}
E_z (\rho, z) = - E_z (\rho, -z); \; \; \; E_{\rho} (\rho, z) = E_{\rho} (\rho, -z),
\label{Esym}
\end{equation}
and
\begin{equation}
 B_z (\rho, z) =  B_z (\rho, -z); \; \; \; B_{\rho} (\rho, z) = - B_{\rho} (\rho, -z).
\label{Bsym}
\end{equation}

We can easily evaluate the behaviour of charge and current distributions far away from the origin. We can take the field of a magnetic
dipole and of concentrated charge as zeroth approximation satisfying Maxwell's equations, then insert them into the right-hand sides of 
eqs. (\ref{Maxstat}) and compute the first corrections, supposing that the fields ${\bf E}$ and ${\bf B}$ develop as:

\begin{equation}
{\bf E} = {\overset{(0)}{\bf E}} + \frac{1}{e^2} {\overset{(1)}{\bf E}} +...., \; \; \; \; 
{\bf B} = {\overset{(0)}{\bf B}} + \frac{1}{e^2} {\overset{(1)}{\bf B}} +....
\label{EBzeroone}
\end{equation}
if we put
\begin{equation}
{\overset{(0)}{\bf E}} = \frac{Q \; \rho}{(\rho^2 + z^2)^{\frac{3}{2}}} {\bf e}_{\rho} + \frac{Q \; z}{(\rho^2 + z^2)^{\frac{3}{2}}} {\bf e}_z,
\label{E-one}
\end{equation}
and 
\begin{equation}
{\overset{(0)}{\bf B}} = \frac{3 \mu \; \rho z}{4 (\rho^2 + z^2)^{\frac{5}{2}}} {\bf e}_{\rho} + \frac{ \mu \;(2 z^2 - \rho^2)}{4 (\rho^2 + z^2)^{\frac{5}{2}}} {\bf e}_z,
\label{B-one}
\end{equation}
where $Q$ is the total charge, $\mu$ the total magnetic moment, then, as first correction, we obtain
\begin{equation}
{\rm div} \; {\overset{(1)}{\bf E}} = \frac{ 3 \mu^2 Q}{8 (\rho^2 + z^2)^{\frac{11}{2}}} (\rho^2 + 10 z^2),
\label{divEone}
\end{equation}
and 
\begin{equation}
{\bf rot} \; {\overset{(1)}{\bf B}} = \frac{ 3 \mu Q^2 \; \rho}{2 (\rho^2 + z^2)^{\frac{9}{2}}} {\bf e}_{\varphi}
\label{rotBone}
\end{equation}
which show that the charge density falls off as $R^{-9}$ and the current density as $R^{-8}$ ($R = \sqrt{\rho^2 + z^2}$), i.e. really fast.
\vskip 0.2cm
The lines of force of the field ${\bf B}$ form a family of closed curves which can be transformed into a family of circles by a suitable
coordinate transformation; the toroidal coordinates are best adapted to describe the situation.
\vskip 0.2cm
Let us introduce the toroidal coordinates $(\mu, \; \eta, \; \phi)$ as:
\begin{equation}
\rho = \frac{a \; \sinh \mu}{\cosh \mu - \cos \eta }, \; \; \; \; z = \frac{ a \; \sin \eta }{\cosh \mu - \cos \eta}, \; \; \phi= \varphi,
\label{toroidal}
\end{equation}
with $0 \leq \phi \leq 2 \pi, \; \; \; 0 \leq \eta \leq 2 \pi $ \; and $0 \leq \mu \leq \infty$; \; $a$ is the constant of dimension of length fixing the scale;
$\mu, \; \eta$ and $\phi$ are dimensionless.  A surface $\mu = \mu_0 =$ Const. is a torus with he external radius $a \; \coth \mu_0$ and internal radius
$a/\sinh \mu_0$. When $\mu \rightarrow \infty$ it reduces to a circle of radius $a$ in the $(x, y)$ plane. When $\mu \rightarrow 0$, the corresponding
circle approaches the $z$-axis.

Toroidal symmetry has been discussed in the context of Skyrme's model (cf.\cite{Meissner},\cite{Kundu}).

The lines of force of ${\bf B}$ coincide with circles $\mu =$ Const., i.e. in new coordinates (\ref{toroidal})
\begin{equation}
{\bf B} = B_{\eta} (\mu, \eta) \; {\bf e}_{\eta}.
\label{B_eta}
\end{equation}
while $B_{\mu} (\mu, \eta) = 0.$
This determines the dependence of ${\bf B}$ on $\eta$:
\begin{equation}
{\rm as} \; \; \; B_{\mu} (\mu, \eta ) = ({\bf rot} {\bf A}) \cdot {\bf e}_{\mu} \; \; \; {\rm with} \; \; \;{\bf A} = A_{\phi} (\mu, \eta) {\bf e}_{\phi},
\label{rotB}
\end{equation}                                                                                 
we have
\begin{equation}
B_{\mu} (\mu, \eta ) = \frac{(\cosh \mu  - \cos \eta )^2}{a \; \sinh \mu} \; \frac{\partial}{\partial \eta} 
\left( \frac{\sinh \mu}{\cosh \mu - \cos \eta} A_{\phi} \right) = 0.
\label{Bmuzero}
\end{equation}
Therefore
\begin{equation}
A_{\phi} (\mu, \eta ) = (\cosh \mu - \cos \eta) \; G(\mu),
\label{Aphi}
\end{equation}
and 
\begin{equation}
B_{\eta} (\mu, \eta) = - \frac{(\cosh \mu - \cos \eta)^2}{a \; \sinh \mu} \frac{\partial}{\partial \eta} \left( \sinh \mu \; G(\mu) \right)
\label{Beta}
\end{equation}
with yet unknown function $G (\mu)$.          

Putting aside for a while the problem of eventual singularity, we can at this point see quite well what the induced charge and current
distributions look like. Consider one of the lines of force of ${\bf B}$, i.e. a circle $\mu = _mu_0, \; \phi = \phi_0$ in the $(\rho, z)$ plane 
(Figure 1, left).                                                                                                                                                                                                                     
The symmetry properties of the fields ${\bf E}$ impose the vanishing of its $\eta$-component for $z=0$, i.e. for $\eta = 0$ or $\pi$, because
$ E_{\eta} (\eta) = -  E_{\eta} (2 \pi - \eta)$. On the other hand,  $B_{\eta} (\eta) = B_{\eta} (2 \pi - \eta)>$, so that the scalar product 
${\bf E} \cdot {\bf B} = E_{\eta} B_{\eta} $ on the circle $\mu = \mu_0$ is an odd function of $\eta$ (Figure 1, right)
\begin{figure}[hbt]
\centering
\includegraphics[width=5.5cm, height=4.5cm]{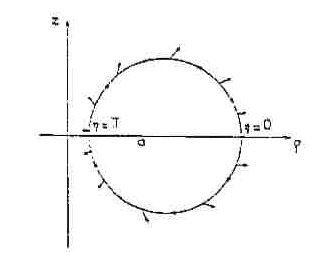} 
\hskip 0.6cm
\includegraphics[width=5.5cm, height=6.5cm]{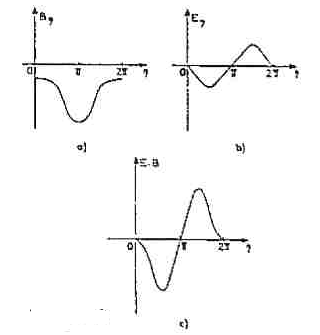} 
\label{fig:E(eta)}
\end{figure}
 
In order to obtain the charge distribution along this circle, we have to compute $- {\bf B} \cdot {\bf grad} ({\bf E} \cdot {\bf B})$,
which reduces to the expression
\begin{equation}
- B_{\eta} \frac{(\cosh \mu - \cos \eta )}{a} \frac{\partial }{\partial \eta} \left( {\bf E} \cdot {\bf B} \right).
\label{chargeexp}
\end{equation}
The corresponding functions are displayed in Figure 3:

\begin{figure}[hbt!]
\centering
\includegraphics[width=8.3cm, height=5cm]{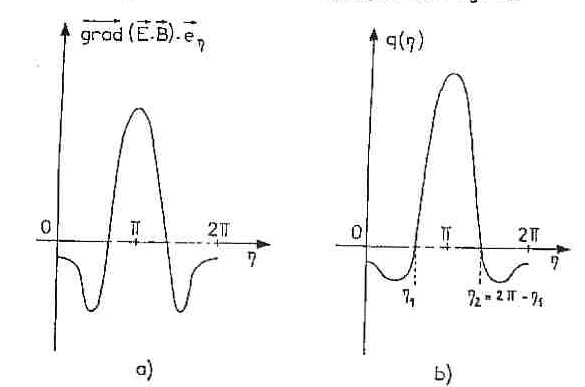} 
\label{fig:Twocurves}
\end{figure}
{\small a) The projection of  ${\bf grad} ({\bf E} \cdot {\bf B})$ on the unit vector ${\bf e}_{\eta}$ as function of $\eta$;
b) The charge density distribution $q(\eta)$ as function of $\eta$. }
\vskip 0.3cm
We see that the charge density changes its sign between $\eta_1$ and $\eta_2 = 2 \pi - \eta_1$. This phenomenon describes
the vacuum polarization: if at the core of the static solution there is an accumulation of charge density of a given sign,
it must be surrounded by a cloud of charge density of opposite sign. The value of $\eta_1$ at which the change of sign occurs depends 
on the line (i.e. the value of $\mu$). Reproducing a similar reasoning for all circles $\mu =$ Const. we obtain the picture of the
overall charge density (Figure 4):
\begin{figure}[hbt!]
\centering
\includegraphics[width=9.2cm, height=5.4cm]{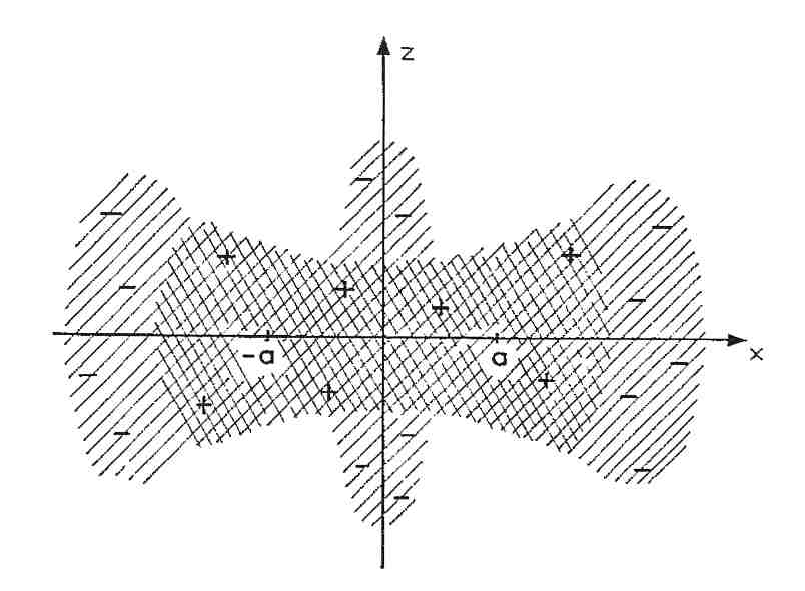} 
\label{fig:Chargedensity}
\end{figure}
{\small The charge density distribution on a cross-section ($x, z, \phi=$ Const.). The strongest vacuum polarization is on the $z$-axis 
and in the symmetry plane $(x,y)$, around the axially symmetric charge distribution at the core.}

If at any point of this distribution we would like to interpret the azimuthal current density obtained from the last equation (\ref{Maxstat})
as being produced by a rotational movement of the charge density around the $z$-axis, then it is easy to see, just comparing
the units (remember that in our equations we have put $c=1$), that the induced charge has to ``move'' with the speed of light.
Of course, nothing is moving here: there is just a distribution of static fields ${\bf E}$ and ${\bf B}$ which produces this illusion,
because the Poynting vector ${\bf E} \times {\bf B}$ has only the azimuthal component everywhere. Nevertheless, the illusion produced
is the same as for the electron as a whole submitted to the {\it ``zitterbewegung''} with the speed $c$ as it comes out from the
Dirac equation. 
\vskip 0.2cm
There is also another striking similarity between the predictions of this model and those of the Dirac equation. The equations (\ref{Maxstat})
are invariant with respect to the independent changes of sign, ${\bf E} \rightarrow - {\bf E}$ and ${\bf B} \rightarrow - {\bf B}$.
This means that any static solution generates automatically three other ones, obtained by the inversions of ${\bf E}$ and ${\bf B}$. Now,
the total charge is linear in ${\bf E}$, while the total magnetic moment is linear in ${\bf B}$; the Poynting vector is proportional
to ${\bf E} \times {\bf B}$, and so will be the total kinetic angular momentum obtained by the integration of ${\bf r} \times ({\bf E} \times {\bf B})$
over entire space. (It is easy to prove that in this case only this {\it orbital contribution} will be different from zero, the internal spin
contribution vanishing when integrated over the space). The total energy remains the same independently of inversions of ${\bf E}$ and ${\bf B}$.

The four solutions so obtained can be put together in the following Table:

\begin{center}
\begin{tabular}{|c|c|c|c|c|}
\hline
\raisebox{0mm}[4mm][2mm] {Solution} & Energy & Charge & Magnetic $\mu$ & Spin  \cr
\hline\hline
\raisebox{0mm}[4mm][2mm]{ ${\bf E}, \; \; \; {\bf B}$ } & m & q & ${\boldsymbol{\mu}}$ & ${\bf S}$ \cr
\hline
\raisebox{0mm}[4mm][2mm]{ $ {\bf E}, - {\bf B}$ } & m & q & $-{\boldsymbol{\mu}}$ & $- {\bf S} $ \cr
\hline
\raisebox{0mm}[4mm][2mm]{ $- {\bf E}, \; \; \;  {\bf B} $ } & m & $-q$ & ${\boldsymbol{\mu}} $ & $-{\bf S}$ \cr
\hline
\raisebox{0mm}[4mm][2mm]{ $-{\bf E}, \; -{\bf B}$ } & m & $-q$ & $- {\boldsymbol{\mu}}$ & ${\bf S}$ \cr
\hline
\end{tabular} 
\end{center}
 
Any static solution is, as a matter of fact, a quadruplet of solutions with the same rest mass. The first two solutions describe a particle
with electric charge $q$ and magnetic moment ${\boldsymbol{\mu}}$ {\it parallel} to  spin ${\bf S}$, in states with spin up or down
(with respect to the $z$-axis); the second pair of solutions describes a particle with the opposite charge $- q$ and magnetic moment
{\it antiparallel} to the spin ${\bf S}$, also in two states with spin up or down. This result is identical with the predictions of 
Dirac's equation for the electron, which leads to the existence of the positron and a half-integer spin, which seems to be a good news.
\vskip 0.2cm
The bad news is that unfortunately no $C^{\infty}$-class solutions of the system (\ref{Maxstat}) exist. The proof is simple and goes as follows:
\vskip 0.2cm
Knowing that ${\rm div} \; {\bf B} = 0$, we can write
\begin{equation}
{\bf B} \cdot {\bf grad} ({\bf E} \cdot {\bf B} ) = {\rm div} \; ({\bf B} \; ({\bf E} \cdot {\bf B} ))
\label{divgrad}
\end{equation} 
Similarly,
\begin{equation}
{\bf E} \times {\bf grad} ({\bf E} \cdot {\bf B}) = {\bf rot} ({\bf E} \; ({\bf E} \cdot {\bf B})),
\label{rotgrad}
\end{equation}
because ${\bf rot} \; {\bf E} = 0$. This leads to
\begin{equation}
{\rm div} \; \left( {\bf E} + \frac{3 }{e^2} \; {\bf B} ({\bf E} \cdot {\bf B}) \right) = 0, \; \; \; \; 
{\bf rot} \; \left( {\bf B} - \frac{3}{e^2} \; {\bf E} ({\bf E} \cdot {\bf B}) \right) = 0.
\label{divErotB}
\end{equation}
If the space we are working in has the topology of $R^3$, and all the functions are supposed to be $C^{\infty}$-smooth, then the Poincar\'e lemma states that
\begin{equation}
{\bf E} + \frac{3 }{e^2} \; {\bf B} ({\bf E} \cdot {\bf B}) = {\bf rot} \; {\bf C} \; \ ; \; \; {\rm and} \; \; \; \; 
{\bf B} - \frac{3}{e^2} \; {\bf E} ({\bf E} \cdot {\bf B}) = {\bf grad} \psi.
\label{Poincare}
\end{equation}
with ${\bf C} ({\bf r})$ and $\psi ({\bf r})$ supposed to be $C^{\infty}$ smooth (vector and scalar, respectively) functions of ${\bf r}$.
Taking the scalar product of the first equation in (\ref{Poincare}) by ${\bf E}$ and of the second equation by ${\bf B}$ we get (supposing that ${\bf E} = - {\bf grad} V$):
\begin{equation}
{\bf E}^2 + \frac{3}{e^2} \; ({\bf E} \cdot {\bf B})^2 = {\bf E} \cdot {\bf rot} {\bf C} = - ({\bf grad} V ) \cdot {\bf rot} {\bf C} =
- {\rm div} \; ( V {\bf rot} {\bf C}),
\label{divVrotC}
\end{equation}
and
\begin{equation}
{\bf B}^2 - \frac{3}{e^2} \; ({\bf E} \cdot {\bf B})^2 = {\bf B} \cdot {\bf grad} \psi = {\rm div} (\psi \; {\bf B} ).
\label{divpsiB}
\end{equation} 
Combining equations (\ref{divVrotC}) and (\ref{divpsiB}) together, we have
\begin{equation}
{\bf E}^2 + {\bf B}^2 = {\rm div} ( \psi {\bf B} - V \; {\bf rot} \; {\bf C} ).
\label{energydiv}
\end{equation} 
If we want the total energy, as well as the total charge, to be finite, then both ${\bf E}$ and ${\bf B}$ must decrease at infinity 
at least as $R^{-2}$, so that the right-hand side of (\ref{energydiv}) must be of the order of $R^{-4}$, which means in turn that
the vector field $\psi \; {\bf B} - V \; {\bf rot} \; {\bf C}$ is decreasing at infinity as $R^{-3}$. Applying Gauss-Ostrogradsky theorem
to a finite $3$-volume $\Omega$ and its $2$-dimensional boundary $\partial \Omega$:
\begin{equation}
\int_{\Omega} {\rm div} (\psi \; {\bf B} - V \; {\bf rot} \; {\bf C} ) \;  d^3 {\bf r} =
\int_{\partial \Omega}  (\psi  \; {\bf B} - V \; {\bf rot} \; {\bf C} )\cdot {\rm d}{\boldsymbol \Sigma},
\label{GaussOstro}
\end{equation}
we see that the integral of ${\bf E}^2 + {\bf B}^2$ over a spherical volume of radius $R$ behaves as $R^{-1}$, i.e. it vanishes when
taken over the whole space. Both expressions ${\bf E}^2$ and ${\bf B}^2$ being positive, this means that ${\bf E} = 0$ and ${\bf B}=0$,
unless the solution is not $C^{\infty}$ and the Poincar\'e lemma does not hold at least on some line or surface.
\vskip 0.2cm
The impossibility of obtaining a $C^{\infty}$ solution with finite energy can be also seen if we try to construct it by applying the method 
of successive approximations in toroidal coordinates.

The whole problem can be reduced down to two equations for two unknown functions, the azimuthal component of the vector potential $A_{\phi}$
and the scalar potential $V$. We can believe that in basic state the dependence on the azimuthal angle $\phi$ is trivial, therefore
we may set
\begin{equation}
A_{\phi} = A_{\phi} (\mu, \eta) \; \; \; {\rm and} \; \; \; V = V (\mu, \eta)
\label{AphiV}
\end{equation}
The dependence of both potentials on the toroidal angle $\eta$ must be of the form $\sin (k \eta)$ or $\cos (k \eta), \; \;  k=1,2,...$;
using the substitution
\begin{equation}
A_{\phi} = u (\eta) \; \sqrt{\cosh \mu - \cos \eta}  = (\cosh \mu - \cos \eta) \; G (\mu)
\label{Aphisubst}
\end{equation}
we make the $\mu$-component of the magnetic field vanish, $B_{\mu} = 0$, and with another substitution
\begin{equation}
V = v (\eta) \; \sqrt{\cosh \mu - \cos \eta} 
\label{Vsubst}
\end{equation}
the laplacians appearing on the left-hand side of equations (\ref{Maxstat}) will have their variables separated.
For example, the equation
\begin{equation}
{\rm div} \; {\bf E} = -  \frac{3 }{e^2} \; {\bf B}\cdot {\bf grad} ({\bf E} \cdot {\bf B})
\label{laplacian}
\end{equation}
will take on the form
\begin{equation}
\frac{1}{\sinh \mu} \frac{\partial}{\partial \mu} \left( \sinh \mu \frac{\partial v}{\partial \mu} \right) +
\frac{\partial^2 v}{\partial \eta^2} + \frac{1}{4} v = 
\frac{3}{a^2 e^2} \frac{(\cosh \mu - \cos \eta)}{\sinh^2 \mu} \left[ \frac{\partial }{\partial \mu}\left( \sinh \mu \; G(\mu)\right) \right]^2
[ W(\mu, \eta) ],
\label{Laplacev}
\end{equation}
with 
\begin{equation}
W = [\cosh \mu - \cos \eta) \frac{\partial^2 v}{\partial \eta^2} + 4 \sin \eta \frac{\partial v}{\partial \eta} 
+ \frac{(5 \sin^2 \eta + 2 \cosh \mu \cos \eta - \cos^2 \eta)}{4 ( \cosh \mu - \cos \eta )}.
\label{Wfunction}
\end{equation}
Similarly, the laplacian of the function $u(\mu, \eta)$ will be equal to some non-linear terms mutiplied by $3/(a^2 e^2$.
\vskip 0.2cm
Developing functions $u$ and $v$ as e.g.  
$\sum_{n=1}^{\infty} \;[ {\overset{(1)}{v}}_n (\mu) \sin (n \eta) +  {\overset{(2)}{v}}_n (\mu) \cos (n \eta) ]$
the second derivatives in (\ref{Laplacev}) will be replaced by $n^2 v$, and the solutions of the {\it homogeneous equations},
which correspond to the zeroth approximation ($\frac{2}{a^2 e^2} = 0$) are given as a series in spherical harmonics 
of half-integer order (cf. Morse and Feshbach, \cite{MorseFeshbach}) 
\begin{equation}
P_{n+ \frac{1}{2}} (\cosh \mu) \; \; \; {\rm and} \; \; \; Q_{n - \frac{1}{2}} (\cosh \mu)
\label{PnQn}
\end{equation}
The functions $P_{n+ \frac{1}{2}}$ display a logarithmic singularity for $\mu = \infty$, i.e. on the circle $\rho = a$,
whereas the functions $Q_{n- \frac{1}{2}}$ have a logarithmic singularity for $\mu = 0$ (i.e. $\rho, z \rightarrow \infty$).
In order to avoid singularity we may use the combination of both, but the price to pay is the discontinuity for some value
of $\mu$ (on some toroidal surface). If we feed in such a solution to the rightèhand side and use the Green functions in
order to compute the first correction, we shall be faced with exactly the same problem, because any Green function has
at least one singularity of the same kind. 

However, the impossibility of producing a non-singular soliton is probably due to the fact that we have projected everything 
onto three space dimensions, discarding the fifth circular one. It seems possible to obtain solitons using the fifth dimension
in a non-trivial way, like in the case of Kaluza-Klein monopoles of Sorkin (\cite{Sorkin}) and Gross and Perry (\cite{GrossPerry}).

Another development should include the non-abelian generalization of the Kaluza-Klein theory into more dimensions, in which
also higher order invariants of the Riemann tensor might be included to the generalized lagrangian.
\vskip 0.3cm
\centerline{ ***************************************************************************************** }
\vskip 0.5cm
\begin{figure}[hbt!]
\centering
\includegraphics[width=7cm, height=8.7cm]{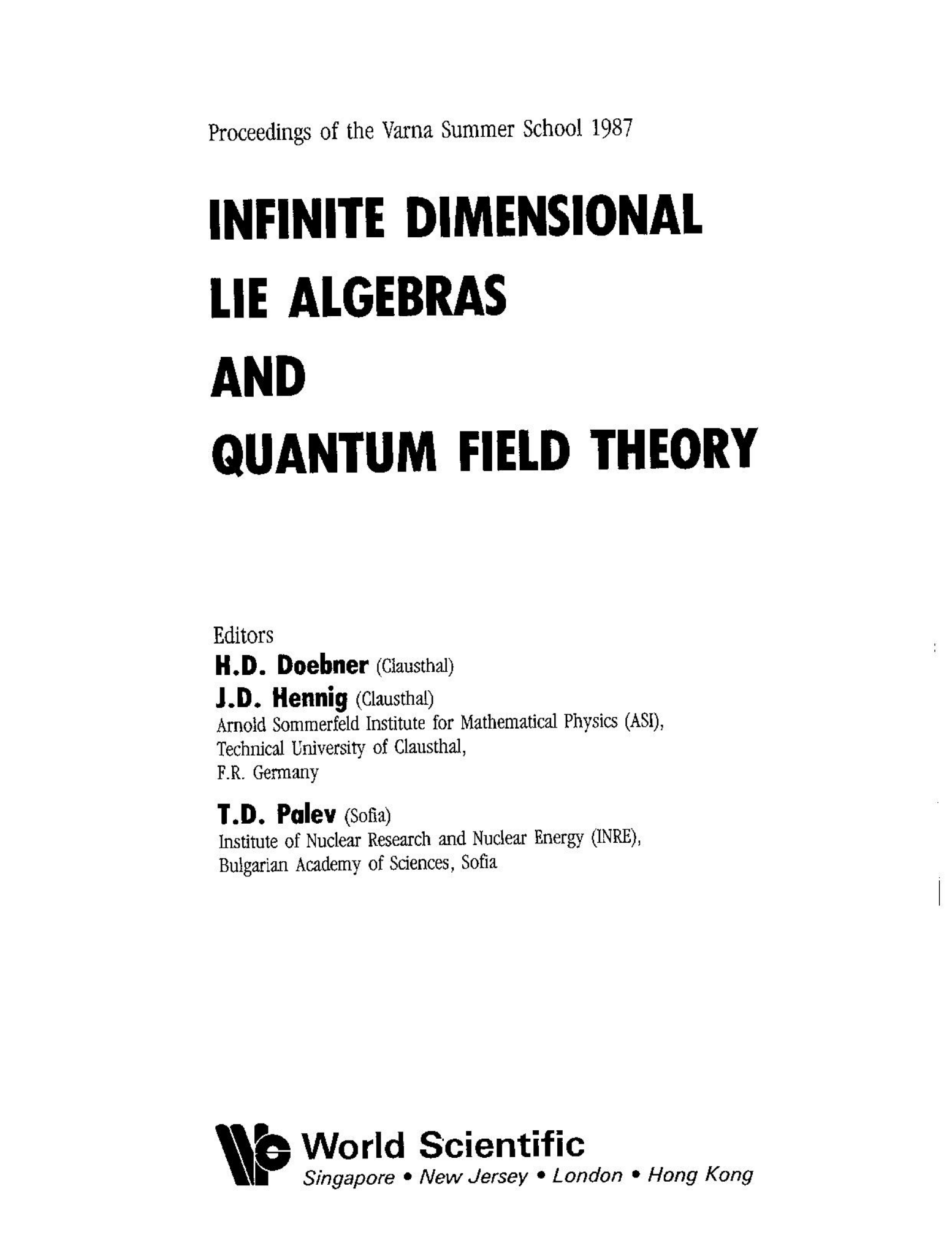} 
\hskip 0.4cm
\includegraphics[width=7cm, height=8.7cm]{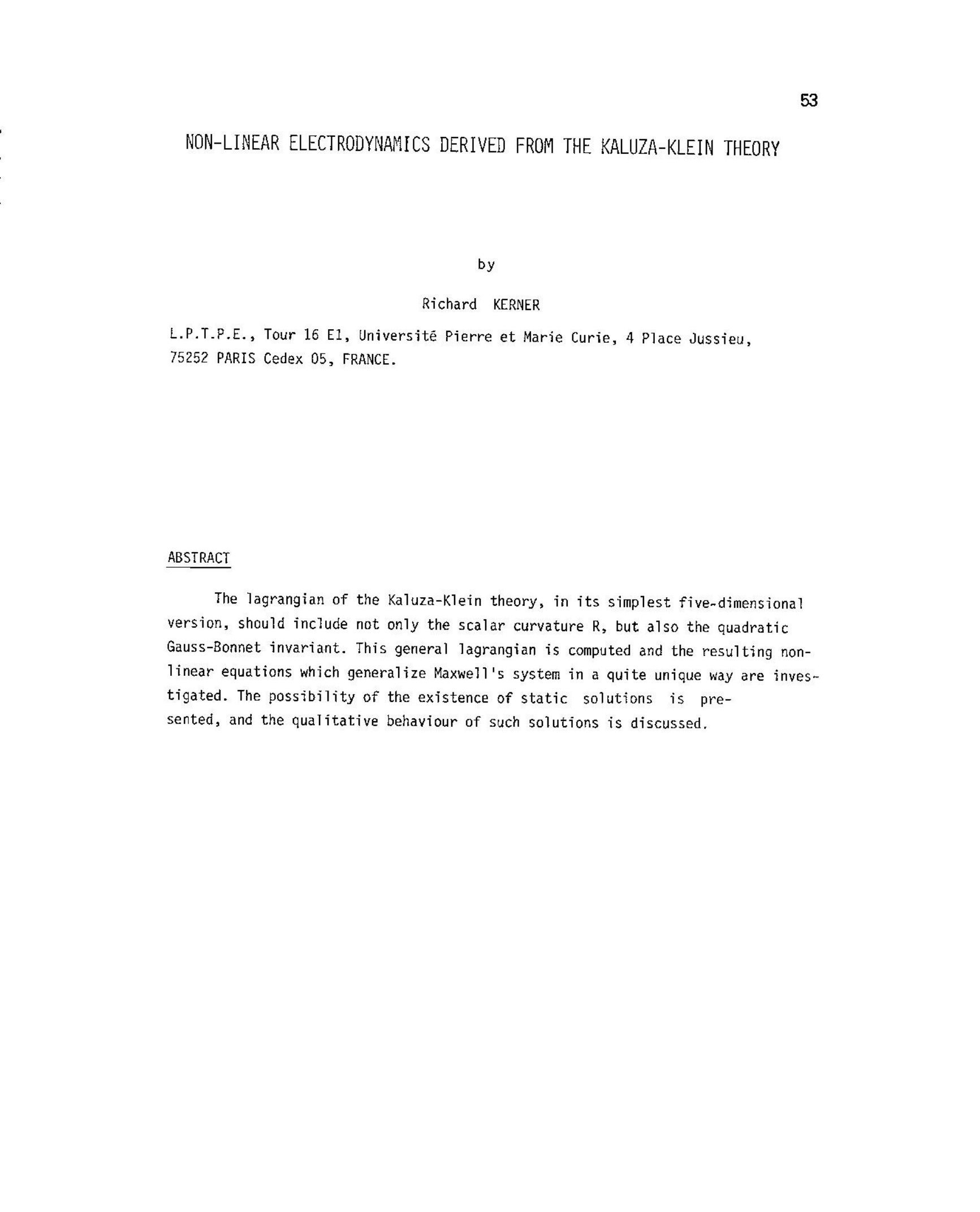} 
\label{fig:Coverpage}
\end{figure}

{\small The cover page of the Proceedings of Varna Summer School Conference of $1987$, ``Infinite Dimensional Lie Algebras and Quantum Field Theory'',
(World Scientific, Editors: H.D. Doebner, J.D. Hennig and T.D. Palev), and the first page of this paper published within, pages $53-72$.}

\newpage

{\large {\bf Acknowledgements}}
\vskip 0.3cm

The enlightening remarks of M. Dubois-Violette are gratefully acknowledged as well as the useful discussions with J. Madore and R. Vinh Mau.

\end{document}